# Nonlinear terahertz devices utilizing semiconducting plasmonic metamaterials

Huseyin R Seren[1], Jingdi Zhang[2,3], George R Keiser[2,5], Scott J Maddox[4], Xiaoguang Zhao[1], Kebin Fan[1], Seth R Bank[4], Xin Zhang[1]*, and Richard D Averitt[2,3]*

[1]Laboratory for Microsystems Technology, Boston University, Dept. of Mech. Eng., Boston, MA, USA
[2]Dept. of Physics, Boston University, Boston, MA, USA
[3]Dept. of Physics, UC San Diego, La Jolla, CA USA
[4]Microelectronics Research Center, The University of Texas at Austin, Austin, TX, USA
[5]School of Engineering, Brown University, Providence, RI, USA

*Correspondence and requests for materials should be addressed to XZ (email: xinz@bu.edu, 8th Saint Mary's Street Photonics Center, 902, Boston, MA 02215, Tel: +1(617)353-0605 Fax: +1(617) 353-5866) or to RDA (email: raveritt@ucsd.edu, UCSD Dept. of Physics, 9500 Gilman Dr., La Jolla, CA 92093, Tel: +1(858) 534-5976 Fax: +1(858) 353-9393).

ABSTRACT

The development of responsive metamaterials has enabled the realization of compact tunable photonic devices capable of manipulating the amplitude, polarization, wave vector, and frequency of light. Integration of semiconductors into the active regions of metallic resonators is a proven approach for creating nonlinear metamaterials through optoelectronic control of the semiconductor carrier density. Metal-free subwavelength resonant semiconductor structures offer an alternative approach to create dynamic metamaterials. We present InAs plasmonic disk arrays as a viable resonant metamaterial at terahertz frequencies. Importantly, InAs plasmonic disks exhibit a strong nonlinear response arising from electric field induced intervalley scattering resulting in a reduced carrier mobility thereby damping the plasmonic response. We demonstrate nonlinear perfect absorbers configured as either optical limiters or saturable absorbers, including flexible nonlinear absorbers achieved by transferring the disks to polyimide films. Nonlinear plasmonic metamaterials show potential for use in ultrafast THz optics and for passive protection of sensitive electromagnetic devices.

The advent of active and tunable metamaterials (MMs) introduced a new path towards controlling light-matter interactions with the possibility to impact photonic applications spanning from microwave to visible frequencies.[1] Nonlinear MMs represent an important class of active electromagnetic composites that can potentially pave the way to produce tailored nonlinear optical phenomena such as harmonic generation or self-focusing.[2] Pioneering prominent examples of nonlinear MMs have been demonstrated in the microwave region where nonlinear lumped circuit elements were used.[3] At infrared frequencies, field enhancement plasmonic MMs provide an important route towards creating enhanced nonlinear composites.[4–7]

Optically responsive materials at THz frequencies have also been demonstrated in the past decade through judicious MM design.[8–10] This includes the creation of dynamically tunable MM devices employing optical, mechanical, or electrical control methods.[1,11–15] Nonlinear THz MMs have also been demonstrated.[16–20] The majority of tunable and nonlinear THz MMs have incorporated semiconductors into the active region of split ring resonators to enable dynamic tuning of the electromagnetic response.[13,14,21,22] It is also possible to exclusively employ semiconductors to create plasmonic devices at THz and infrared frequencies.[23–25] The plasma frequency of semiconductors can be tuned by adjusting the doping level, providing a path towards THz plasmonic semiconductor metamaterials (PSMM).[26–29] The response of PSMM can be tailored via structure and geometry and can be modulated using, as examples, electric, magnetic, and thermal stimuli. Importantly, semiconductors exhibit large nonlinearities at THz frequencies [30–36] enabling (as demonstrated below) nonlinear plasmonics and providing a key capability for future terahertz circuits and systems.[6,37]

We have created PSMMs using n-doped InAs (n-InAs) thin films patterned into disk arrays that are resonant at THz frequencies. As is well known, the plasmonic response of particle arises from dielectric confinement. This results in a resonant response that is determined by the geometry and the carrier concentration with the quality of the resonance determined by the scattering rate or, in the case of semiconducting plasmonic particles, the mobility. Thus, n-InAs is an attractive plasmonic material at THz frequencies because of the high mobility (~20×10$^3$ cm$^2$/V-s) and ability to control the resonance frequency through doping. Further, as we demonstrate, InAs disks exhibit a strong nonlinear response. In particular, high-field terahertz nonlinear transmission measurements reveal that the disk plasmon resonance exhibits a nonlinear response arising from field induced intervalley scattering of conduction band electrons to a low-mobility satellite valley. We utilized these nonlinear PSMMs to create both saturable absorbers and optical limiters by incorporating a ground plane to create a perfect absorber geometry including flexible absorbers created via transfer of the InAs arrays to polyimide.

A PSMM composed of 70 μm diameter n-InAs disks with 100 μm hexagonal lattice periodicity was fabricated as shown in Figure 1a, and b. The geometry was formed by dry etching of a 2μm thick n-InAs film grown on SI-GaAs using molecular beam epitaxy.[38] Our samples were doped to 10$^{17}$ cm$^{-3}$, as shown in Figure 2, to obtain a strong plasmonic response at ~0.8THz. The band structure, depicted in **Figure 1**c plays an important role in the plasmonic response. At low electric fields, the free electrons reside predominantly in the Γ-valley and exhibit a small effective mass and high mobility. The oscillator strength of the plasmon resonance can, in principle, be modified at high electric fields through intervalley scattering or impact ionization[30–35] (Figure 1c). For example, efficient intervalley scattering (Γ→ L) would result in a damping of the plasmon resonance because of the considerably larger effective mass and reduced mobility of carriers in the L-valley.

High field THz time domain spectroscopy (TDS) was employed to characterize the samples utilizing tilted pulse front generation in lithium niobate.[39,40] Figure 2a shows the measured PSMM plasmon resonance as a function of frequency at various incident field strengths with an estimated maximum around $E_0 \approx 300$ kV/cm. At the lowest field ($0.1E_0$, the black curve in Figure 2a), a plasmon resonance is evident at 0.77 THz with a transmission of approximately 35%. With increasing field strength there is an increase in the transmission associated with plasmon damping. At the highest field strength ($E_0$, the purple curve in Figure 2a) the transmission has increased to approximately 75%, an increase of 40% in comparison to the low field case. Figure 2c shows the corresponding phase as a function of frequency (relative to the phase at the highest field strength). The largest phase shift (between the lowest and highest fields) at ~0.9 THz was 35°; at 1.2 THz (well above resonance) it has decreased to ~22°. The observed damping is consistent with THz electric field induced intervalley scattering (Γ→ L) resulting in a decrease in the average electron mobility.[41]

To gain insight into the nonlinear response of the InAs disks, we modeled the PSMMs using CST Microwave Studio utilizing the Drude model to describe the electromagnetic response of InAs (see supplementary information for details). In agreement with experiment, the simulated resonant response at ~0.77 THz corresponds to the dipolar plasmonic mode of the disk structure. To simulate the increased carrier scattering we decreased the electron mobility, which results in a quenching of the plasmon oscillator strength in agreement with experiment (Figure 2b). This interpretation is consistent with intervalley scattering leading to coexisting populations of electrons in the Γ and L valleys which yields an average effective mobility as indicated in Figure 2d. Specifically, our simulations are consistent with a decrease in the effective mobility from $19\times10^3$ cm$^2$/V-s to $3.5\times10^3$ cm$^2$/V-s upon increasing the field from $0.1E_0$ to $E_0$. Ignoring effects due to nonparabolicity, impact ionization, and scattering to the X-valley, and assuming the Γ and L-valleys have constant mobilities of $20\times10^3$ and 20 cm$^2$/V-s, respectively, the observed change is consistent with greater than 50% of the carriers being transferred to the L-valley. In the experimental data, there is also a slight shift in the resonance frequency; we attribute this shift to the decrease in the plasma frequency caused by an increase in effective mass arising from band nonparabolicity and transfer to the L-valley at high fields.[42] This effect is well captured in the simulations by increasing the effective electron mass ($m_{eff}$ = 0.023 → 0.035) in the Drude model.

The ability to fabricate metamaterials from semiconductor resonators opens up vast opportunities for creating nonlinear active devices. In the following we demonstrate nonlinear absorbers (i.e. saturable absorbers and optical limiters). Similar devices operating in the infrared currently find practical use in ultrafast optics, mode locking, and sensor/eye protection.[43,44] Indeed, bulk semiconductors can show a nonlinear absorption of THz light due to several electronic nonlinear processes.[45–47] However, using semiconductors in a MM perfect absorber geometry[48] provides additional control over the nonlinearity, absorption strength, frequency, and modulation depth reducing the device thickness in comparison to an unpatterned semiconductor device.

Perfect absorption phenomenon in MM absorbers can be explained in terms of impedance matching of metamaterial effective parameters[8] or in terms of interference of the reflected light from resonator and ground plane layers.[49] The highest absorbance is achieved over a narrow range of optimized conditions, with the absorbance becoming weaker (or stronger) as the MM properties are modified. Thus, as the properties of the PSMM layer are altered due to nonlinearities the optimal absorbance conditions change. PSMM absorbers can be designed such that the highest absorbance occurs at either low fields or at high fields, resulting in saturable absorbers or optical limiters, respectively.

PSMMs in a hexagonal array (70 μm diameter and 90 μm periodicity) were made into absorbers by spin coating a polyimide layer followed by physical evaporation of a gold ground plane as additional steps to PSMM fabrication (**Figure 3**a and b). The PSMM geometry was chosen to obtain a resonance at ~0.5 THz, which is close to the peak of our THz high-field source, thereby providing the largest possible THz field for this nonlinear proof-of-principle demonstration. Two optimized thicknesses were used for the polyimide spacer layer; for the saturable absorber and optical limiter, the thickness of the polyimide spacer layer was 18 μm and 40 μm, respectively. The saturable absorber (SA) is designed to maximize the absorption at low field strengths by taking advantage of the high carrier mobility n-InAs. The optical limiter (OL) is optimized to increase the absorption at higher field strengths where the effective electron mobility is decreased. Importantly, these devices are thinner than a quarter wavelength at the peak absorption frequency of 0.46 THz (e.g. $\lambda/4 \approx 160$ μm in free space and 45 μm in GaAs).

The ground plane prevents illumination from the front side and has zero transmission. Thus, the reflection from the absorbers was measured from the bare GaAs side of the device at near-normal incidence. This is shown schematically in Figure 3c. The absorber acts as a Fabry-Perot etalon and produces multiple time resolved THz pulses in the TDS reflection signal. The first reflected pulse is directly from the front surface of the substrate and contains no information about the PSMM absorption.

To determine the absorption of the active layer of the PSMM absorber (i.e. the *internal absorbance*) we windowed the data to isolate second reflected pulse, which corresponds to the first reflection from the device layer (see supplementary information). As a reference, a sample composed of a SI-GaAs substrate, polyimide spacer and the ground plane was used. Utilizing the second reflected pulses from both the device and the reference provides a normalized measure of the internal absorbance. We followed the same procedure for the full wave electromagnetic simulations using a time-domain transient solver.

The measurement and simulation results are shown in Figure 3d-g for the SA and OL absorbers. The absorbance for each device was calculated by subtracting the reflectance from unity ($A = 1 - |E_{reflected}|^2$), since the transmission through the ground plane was negligible. For the SA, the absorbance at the resonant frequency of 0.5 THz was 97.5% at low fields ($0.2E_0$, the black curve in **Figure 3**d). With increasing field strength, the absorbance monotonically decreases. At the full strength of the THz beam ($E_0$, the purple curve in Figure 3d), the absorbance was reduced to 49%. Thus, between $0.2 - 1.0\ E_0$, a modulation of nearly 50% was obtained. The inset of Figure 3d shows the change in the absorbance as a function of field strength.

The measurement results for the optical limiter are shown in **Figure 3**f. The absorbance was 80% at low fields ($0.2E_0$, the black curve in Figure 3f). As the field strength is increased ($0.5E_0$, the blue curve in Figure 3f), the absorbance increased to ~99%. Further increasing the field strength up to $E_0$, the absorbance dropped back to 80%, indicating that the optimal OL effect for this particular device occurred at $0.5E_0$. The inset of Figure 3f clearly shows the nonlinear modulation of the absorbance at 0.46 THz. Another consequence of using a thicker spacer layer in the optical limiter was the observed broadband absorption. This is mainly because of a second absorption mode arising around 1 THz whose tail merged with the fundamental absorption mode. While the modulation was much smaller for the OL in comparison to the SA, these results nonetheless reveal that it is possible to create both SAs and OLs from nonlinear MMs.

As in Figure 2, the nonlinear response of the PSMM absorbers were modeled using the Drude response of n-InAs, with both the mobility and effective mass used as variable parameters (see supplementary information Table S1). The overall trends in the simulated absorbances (Figure 3e, g) are in good agreement with the measurements. The origin of the differences in resonances between experiment and simulation arise from several sources. In particular, for the high-field THz reflectance measurements, it is difficult to remove artifacts arising from scattering, and slight deviations in the fabricated structures were not captured in the simulations.

While the PSMM absorbers with a GaAs substrate demonstrate the feasibility of creating MM SAs and OLs, it is desirable to eliminate the substrate to achieve real perfect absorption (i.e. to eliminate the first surface reflection from the substrate). Transfer printing techniques have recently been developed making it possible to transfer semiconductors onto flexible materials with the aid of a sacrificial layer.[50] We have fabricated substrate-free PSMM absorbers using n-InAs film grown on an AlAsSb sacrificial layer grown via MBE on a SI-GaAs substrate (**Figure 4**a-c). To facilitate etching based lift-off of the InAs, we used ring shaped resonators with holes that extend through the ground plane creating a perforated sheet. Although this perforated structure is not optimal for perfect absorbers, the diameter of the holes is much smaller than the resonance wavelength and the transmission through them is negligibly small. Moreover, tests were conducted by placing the samples on gold mirrors, ensuring zero transmission. A gold mirror was also used as a reference.

This substrate-free device was optimized to act as a SA; it exhibited an 83% resonant absorbance peak at $0.15E_0$ and ~0.6 THz (the black curve in **Figure 4**d). The absorbance dropped to 58% at $0.5E_0$ field strength (the green curve in Figure 4d). The maximum absorption did not reach unity due to fabrication imperfections (e.g. the fabricated polyimide thickness came out 4 μm thinner than the designed thickness of 29 μm). Further, defects were formed on the InAs film during the transfer printing which potentially decreased the carrier mobility. Nonetheless, absorption saturation of ~25% was successfully demonstrated. The simulation results are shown in Figure 4e (see supplementary information Table S2 for the model). The dashed line in Figure 4e shows the designed maximum absorbance available for a thicker, better optimized polyimide layer and for the expected mobility of a defect-free InAs film.

One of the design concerns of MM absorbers is the dependence of the absorption on the incidence angle ($\theta$) and incident polarization ($\phi$). Due to complications with the high-field optical setup for oblique angles, we investigated these scenarios for the substrate-free PSMM absorber via simulation. The simulated $\theta$ and $\phi$ dependence of the substrate-free PSMM absorber are shown in Figure 4f. For TE polarized light, there was a slight absorption degradation for $\theta>45$ degrees, while for TM polarized light, absorption was maintained over a wide range of angles with a slight shift to higher frequencies.

Our results suggest that PSMMs are of potential use in applications that include ultrafast THz optics and as protective layers from strong resonant electromagnetic fields. The ability to create flexible nonlinear devices further facilitates applications since these PSMMs can conformally adhere to curved surfaces. Plasmonic semiconductor resonators complement existing metallic structures, such as split ring resonators, providing alternative fabrication strategies to create active materials with reduced local field enhancement and correspondingly higher damage thresholds. Geometries other than disk and ring arrays (e.g. dimers or bowties) can also be employed to obtain useful functionality. Finally, the doping level and defect density in InAs can be engineered to control the resonance frequency, strength, and response time of MM devices. In principle, even higher mobility materials (e.g. InSb, or two-dimensional electron gases) will yield even sharper plasmonic resonances.

**Methods:**

*Fabrication:* For our studies, 2 μm thick n-InAs films were grown via molecular beam epitaxy (MBE) with a Si doping concentration of $10^{17}$ cm$^{-3}$ on a 500μm thick semi-insulating SI-GaAs substrate. A 100 nm thick Ti mask layer was patterned on the film using a lift-off technique. Next, an InAs film on a 1×1 cm$^2$ die was etched thoroughly using reactive ion etching with a gas mixture of H$_2$, CH$_4$, and Ar. The Ti mask was then etched away in HF solution. For the perfect absorbers on GaAs substrates, this procedure was followed by polyimide spin coating and curing, and 150 nm gold evaporation. For the fabrication details of the PSMM absorber on flexible substrate, see SI text.

*Terahertz Time Domain Spectroscopy:* The THz-TDS setup makes use of the tilted-pulse-front technique to generate THz pulses from a LiNbO$_3$ crystal (see SI-Figure 1). The THz pulses used in this experiment were ~1ps in duration with a maximum electric field strength of ~300 kV/cm. The field strength incident on the sample is controlled through a pair of linear polarizers. Example time domain measurements of the PSMM and SI-GaAs reference are presented in SI Figure 2.

*Simulations:* In our simulations, we used the Drude response for n-InAs with the following parameters: $N_d$ = 1e17 cm$^{-3}$, $\varepsilon_\infty$ = 12.25, $m_{eff}$ = 0.023, and $\mu$ = 3.5e3 - 1.9e4 cm2/V-s. The SI-GaAs substrate is assumed to have a relative dielectric constant of 12.94 with a frequency independent loss tangent of 0.006. A unit cell and a reference sample were simulated in the time domain and the results were obtained in the same fashion as in the experiments (e.g. by Fourier transforming the time-domain data and performing a Fresnel analysis in the frequency domain). To include nonlinearities caused by intervalley scattering, we used a variable mobility and effective mass in our Drude model of InAs. By varying the relationship between mobility and effective mass, we were able to account for the field dependent collision frequency. For further detail see SI text.

**References:**

1. Zheludev, N. I. & Kivshar, Y. S. From metamaterials to metadevices. *Nat. Mater.* **11,** 917–924 (2012).

2. Lapine, M., Shadrivov, I. V. & Kivshar, Y. S. Colloquium: Nonlinear metamaterials. *Rev. Mod. Phys.* **86,** 1093–1123 (2014).

3. Shadrivov, I. V., Kozyrev, A. B., Van Der Weide, D. W. & Kivshar, Y. S. Tunable transmission and harmonic generation in nonlinear metamaterials. *Appl. Phys. Lett.* **93,** (2008).

4. Jain, P. K., Xiao, Y., Walsworth, R. & Cohen, A. E. Surface plasmon resonance enhanced magneto-optics (SuPREMO): Faraday rotation enhancement in gold-coated iron oxide nanocrystals. *Nano Lett.* **9,** 1644–50 (2009).

5. Minovich, A. *et al.* Liquid crystal based nonlinear fishnet metamaterials. *Appl. Phys. Lett.* **100,** (2012).

6. Kauranen, M. & Zayats, A. Nonlinear plasmonics. *Nat. Photonics* **6,** 737–748 (2012).

7. Nikolaenko, A. E. *et al.* Nonlinear graphene metamaterial. *Appl. Phys. Lett.* **100,** 2–5 (2012).

8. Landy, N. I., Sajuyigbe, S., Mock, J. J., Smith, D. R. & Padilla, W. J. Perfect metamaterial absorber. *Phys. Rev. Lett.* **100,** 207402 (2008).

9. Tao, H. *et al.* Highly flexible wide angle of incidence terahertz metamaterial absorber: Design, fabrication, and characterization. *Phys. Rev. B Condens. Matter Mater. Phys.* **78,** 241103R (2008).

10. Fan, K., Strikwerda, A. C., Tao, H., Zhang, X. & Averitt, R. D. Stand-up magnetic metamaterials at terahertz frequencies. *Opt. Express* **19,** 12619–12627 (2011).

11. Keiser, G. R., Fan, K., Zhang, X. & Averitt, R. D. Towards Dynamic, Tunable, and Nonlinear Metamaterials via Near Field Interactions: A Review. *J. Infrared, Millimeter, Terahertz Waves* **34,** 709–723 (2013).

12. Zhu, W. M. *et al.* Microelectromechanical Maltese-cross metamaterial with tunable terahertz anisotropy. *Nat. Commun.* **3,** 1274 (2012).

13. Chen, H.-T. *et al.* A metamaterial solid-state terahertz phase modulator. *Nat. Photonics* **3,** 148–151 (2009).

14. Seren, H. R. *et al.* Optically Modulated Multiband Terahertz Perfect Absorber. *Adv. Opt. Mater.* **2,** 1221–1226 (2014).

15. Zhang, Y. *et al.* Gbps Terahertz External Modulator Based on a Composite Metamaterial with a Double-Channel Heterostructure. *Nano Lett.* 150429124146004 (2015). doi:10.1021/acs.nanolett.5b00869

16. Fan, K. *et al.* Nonlinear terahertz metamaterials via field-enhanced carrier dynamics in GaAs. *Phys. Rev. Lett.* **110,** 217404 (2013).

17. Zharov, A. A., Shadrivov, I. V & Kivshar, Y. S. Nonlinear properties of left-handed metamaterials. *Phys. Rev. Lett.* **91,** 37401 (2003).

18. Rose, A., Huang, D. & Smith, D. R. Controlling the second harmonic in a phase-matched negative-index metamaterial. *Phys. Rev. Lett.* **107,** 63902 (2011).

19. Wang, B., Zhou, J., Koschny, T. & Soukoulis, C. C. M. Nonlinear properties of split-ring resonators. *Opt. Express* **16,** 16058–16063 (2008).

20. Liu, M. *et al.* Terahertz-field-induced insulator-to-metal transition in vanadium dioxide metamaterial. *Nature* **487,** 345–348 (2012).

21. Chen, H.-T. *et al.* Experimental demonstration of frequency-agile terahertz metamaterials. *Nat. Photonics* **2,** 295–298 (2008).

22. Fan, K., Strikwerda, A. C., Zhang, X. & Averitt, R. D. Three-dimensional broadband tunable terahertz metamaterials. *Phys. Rev. B* **87,** 161104 (2013).

23. Maier, S. A. *Plasmonics: fundamentals and applications*. (Springer, 2007).

24. Boltasseva, A. & Atwater, H. Low-loss plasmonic metamaterials. *Science* **331,** 290–291 (2011).

25. N'Tsame Guilengui, V., Cerutti, L., Rodriguez, J.-B., Tournié, E. & Taliercio, T. Localized surface plasmon resonances in highly doped semiconductors nanostructures. *Appl. Phys. Lett.* **101,** 161113 (2012).

26. Sensale-Rodriguez, B. *et al.* Broadband graphene terahertz modulators enabled by intraband transitions. *Nat. Commun.* **3,** 780 (2012).

27. Hanham, S. M. *et al.* Broadband terahertz plasmonic response of touching InSb disks. *Adv. Mater.* **24,** OP226–OP230 (2012).

28. Bai, Q. *et al.* Tunable slow light in semiconductor metamaterial in a broad terahertz regime. *J. Appl. Phys.* **107,** 093104 (2010).

29. Hoffman, A. J. *et al.* Negative refraction in semiconductor metamaterials. *Nat. Mater.* **6,** 946–50 (2007).

30. Ho, I.-C. & Zhang, X.-C. Driving intervalley scattering and impact ionization in InAs with intense terahertz pulses. *Appl. Phys. Lett.* **98,** 241908 (2011).

31. Gaal, P. *et al.* Nonlinear Terahertz Response of n-Type GaAs. *Phys. Rev. Lett.* **96,** 187402 (2006).

32. Kuehn, W. *et al.* Terahertz-induced interband tunneling of electrons in GaAs. *Phys. Rev. B* **82,** 075204 (2010).

33. Razzari, L. *et al.* Nonlinear ultrafast modulation of the optical absorption of intense few-cycle terahertz pulses in n-doped semiconductors. *Phys. Rev. B* **79,** 193204 (2009).

34. Su, F. H. *et al.* Terahertz pulse induced intervalley scattering in photoexcited GaAs. *Opt. Express* **17,** 9620–9 (2009).

35. Hebling, J., Hoffmann, M. C., Hwang, H. Y., Yeh, K.-L. & Nelson, K. a. Observation of nonequilibrium carrier distribution in Ge, Si, and GaAs by terahertz pump–terahertz probe measurements. *Phys. Rev. B* **81,** 035201 (2010).

36. Hoffmann, M., Hebling, J. & Hwang, H. THz-pump/THz-probe spectroscopy of semiconductors at high field strengths. *JOSA B* **26,** 29–34 (2009).

37. Boyd, R. W. *Nonlinear Optics*. *Appl. Opt.* **5,** (2003).

38. Trampert, A., Tournié, E. & Ploog, K. H. Novel plastic strain-relaxation mode in highly mismatched III-V layers induced by two-dimensional epitaxial growth. *Appl. Phys. Lett.* **66,** 2265–2267 (1995).

39. Hebling, J., Almasi, G., Kozma, I. & Kuhl, J. Velocity matching by pulse front tilting for large area THz-pulse generation. *Opt. Express* **10,** 1161–1166 (2002).

40. Hirori, H., Doi, a., Blanchard, F. & Tanaka, K. Single-cycle terahertz pulses with amplitudes exceeding 1 MV/cm generated by optical rectification in LiNbO3. *Appl. Phys. Lett.* **98,** 2011–2014 (2011).

41. Arabshahi, H., Khalvati, M. R. & Rokn-Abadi, M. R. Temperature and doping dependencies of electron mobility in InAs, AlAs and AlGaAs at high electric field application. *Brazilian J. Phys.* **38,** 293–296 (2008).

42. Li, Y. B. *et al.* Infrared reflection and transmission of undoped and Si-doped InAs grown on GaAs by molecular beam epitaxy. *Semicond. Sci. Technol.* **8,** 101–111 (1999).

43. Keller, U. & Weingarten, K. Semiconductor saturable absorber mirrors (SESAM's) for femtosecond to nanosecond pulse generation in solid-state lasers. *Sel. Top. Quantum Electron. IEEE J.* **2,** 435–453 (1996).

44. Tutt, L. & Boggess, T. A review of optical limiting mechanisms and devices using organics, fullerenes, semiconductors and other materials. *Prog. quantum Electron.* **17,** 299–338 (1993).

45. Hoffmann, M. C. & Turchinovich, D. Semiconductor saturable absorbers for ultrafast terahertz signals. *Appl. Phys. Lett.* **96,** 151110 (2010).

46. Cao, J. C. Interband Impact Ionization and Nonlinear Absorption of Terahertz Radiation in Semiconductor Heterostructures. *Phys. Rev. Lett.* **91,** 237401 (2003).

47. Tanaka, K., Hirori, H. & Nagai, M. THz Nonlinear Spectroscopy of Solids. *IEEE Trans. Terahertz Sci. Technol.* **1,** 301–312 (2011).

48. Watts, C. M., Liu, X. & Padilla, W. J. Metamaterial electromagnetic wave absorbers. *Adv. Mater.* **24,** OP98–120, OP181 (2012).

49. Chen, H.-T. Interference theory of metamaterial perfect absorbers. *Opt. Express* **20,** 7165–72 (2012).

50. Carlson, A., Bowen, A. M., Huang, Y., Nuzzo, R. G. & Rogers, J. A. Transfer printing techniques for materials assembly and micro/nanodevice fabrication. *Adv. Mater.* **24,** 5284–5318 (2012).

**Acknowledgements:** Work at BU was supported in part by the National Science Foundation under contract ECCS 1309835, and the Air Force Office of Scientific Research under contract FA9550-09-1-0708. JZ and RDA acknowledge support from DOE - Basic Energy Sciences under Grant No. DE-FG02-09ER46643, under which the THz measurements were performed. Work at UT-Austin was supported by a Multidisciplinary University Research Initiative from the Air Force Office of Scientific Research (AFOSR MURI Award No. FA9550-12- 1-0488). The authors also thank Boston University Photonics Center for technical support.

**Author Contributions:** R.D.A., X.Z., S.R.B and H.R.S. developed the idea. S.R.B and S.J.M. performed InAs material growth. H.R.S. and G.R.K. performed electromagnetic simulations and optimizations. H.R.S, K.F., and X.Z. fabricated the device. J.Z. built the measurement setup and conducted the measurements. R.D.A, X.Z. and, S.R.B. supervised the project. H.R.S., G.R.K., X.Z. and R.D.A. wrote the paper. All authors contributed to understanding of the physics and revised the paper.

**Competing financial interests:** The authors declare no competing financial interests.

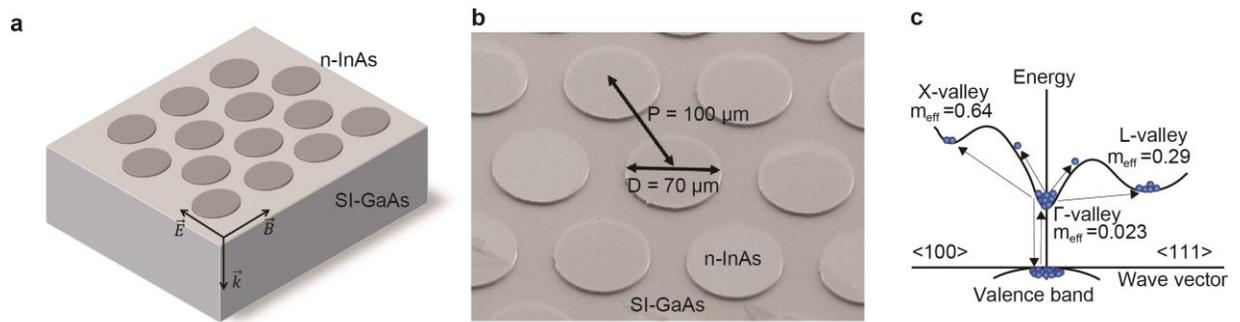

**Figure 1 | Nonlinear Plasmonic Semiconductor Metamaterial. a,** Schematic view of InAs disk array on semi-insulating GaAs. **b,** SEM image of the fabricated PSMM: InAs film thickness: 2 μm, SI-GaAs substrate thickness = 500 μm, disk diameter (D) = 70 μm, periodicity (P) = 100 μm. **c,** Band diagram of InAs showing potential inter- and intra-band transitions triggered by high THz fields (e.g. ballistic acceleration, impact ionization, and intervalley scattering).

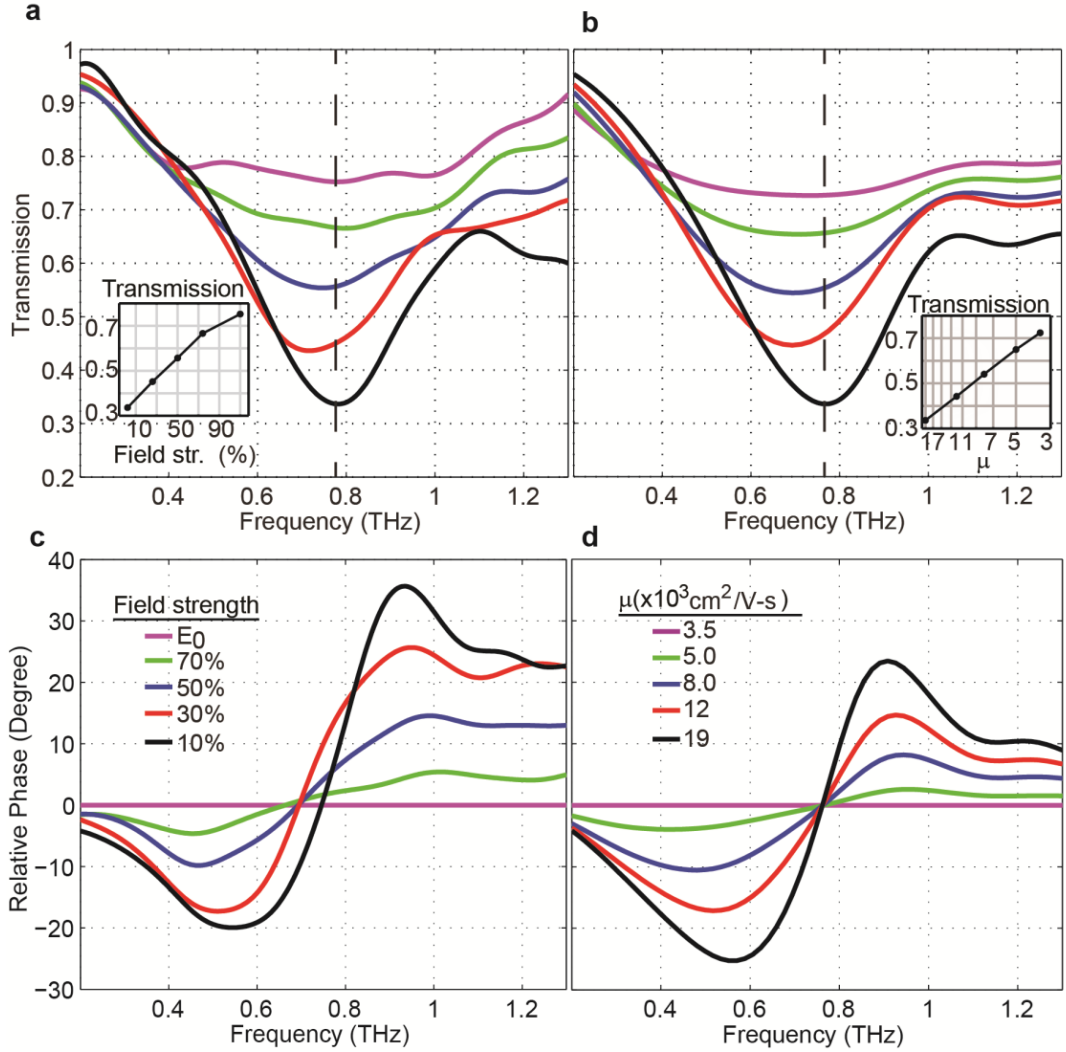

**Figure 2 | Transmisson spectra of the nonlinear PSMM. a,** Measured transmission amplitude of PSMM shown in Figure 1 for various THz field strengths. Inset shows the change in transmission as a function of field strength at the resonance frequency of 0.77 THz. **b**, Simulated transmission amplitude as a function of InAs mobility with $N_d = 1\times10^{17}$ cm$^{-3}$, $\mu$ = 3.5e3 – 1.9e4 cm$^2$/V-s, $\varepsilon_\infty$ = 12.25, $m_{eff}$ = 0.023-0.035. Inset shows the change in transmission as a function of InAs mobility (unit: ×10$^3$ cm$^2$/V-s) at the resonance frequency of 0.77 THz. **c** and **d,** Corresponding measured and simulated transmission phases normalized with respect to the phase at the highest field strength and the lowest electron mobility.

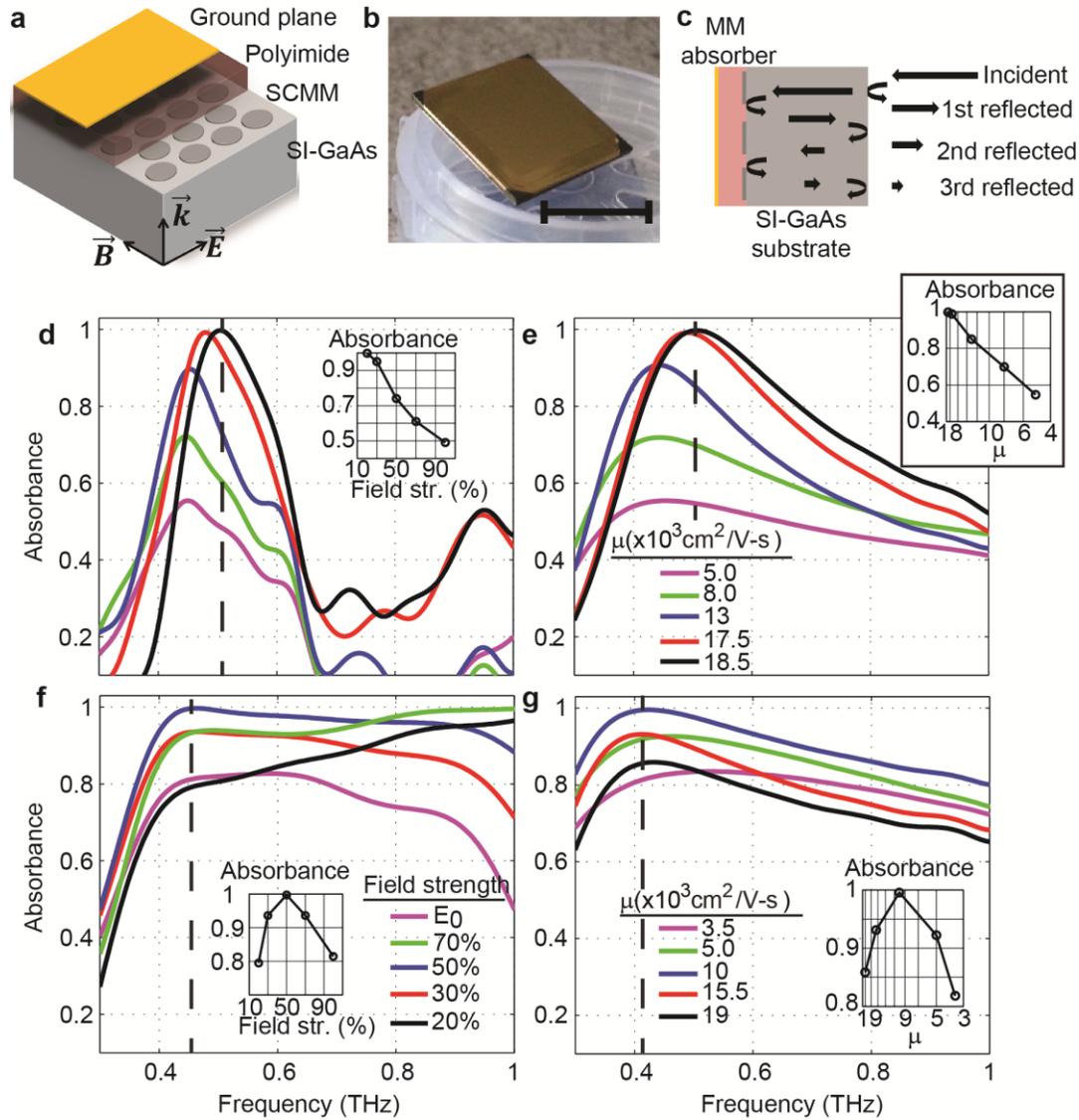

**Figure 3 | Nonlinear PSMM absorbers. a,** Schematic view of the PSMM absorber layers. **b**, A fabricated PSMM device view from the ground plane side with 1 cm$^2$ active area (scale bar is 1 cm). **c**, Schematic representation of the etalon reflections in the GaAs substrate. Measurements and corresponding simulation results for **d - e,** the SA with 18 μm thick polyimide layer, and **f - g,** OL with 40 μm thick polyimide layer, respectively. **d** and **f** share the same legend. Insets show the absorbance trends as a function of field strength and electron mobility (μ) at the frequencies indicated by the dashed lines (mobility unit: ×10$^3$ cm$^2$/V-s).

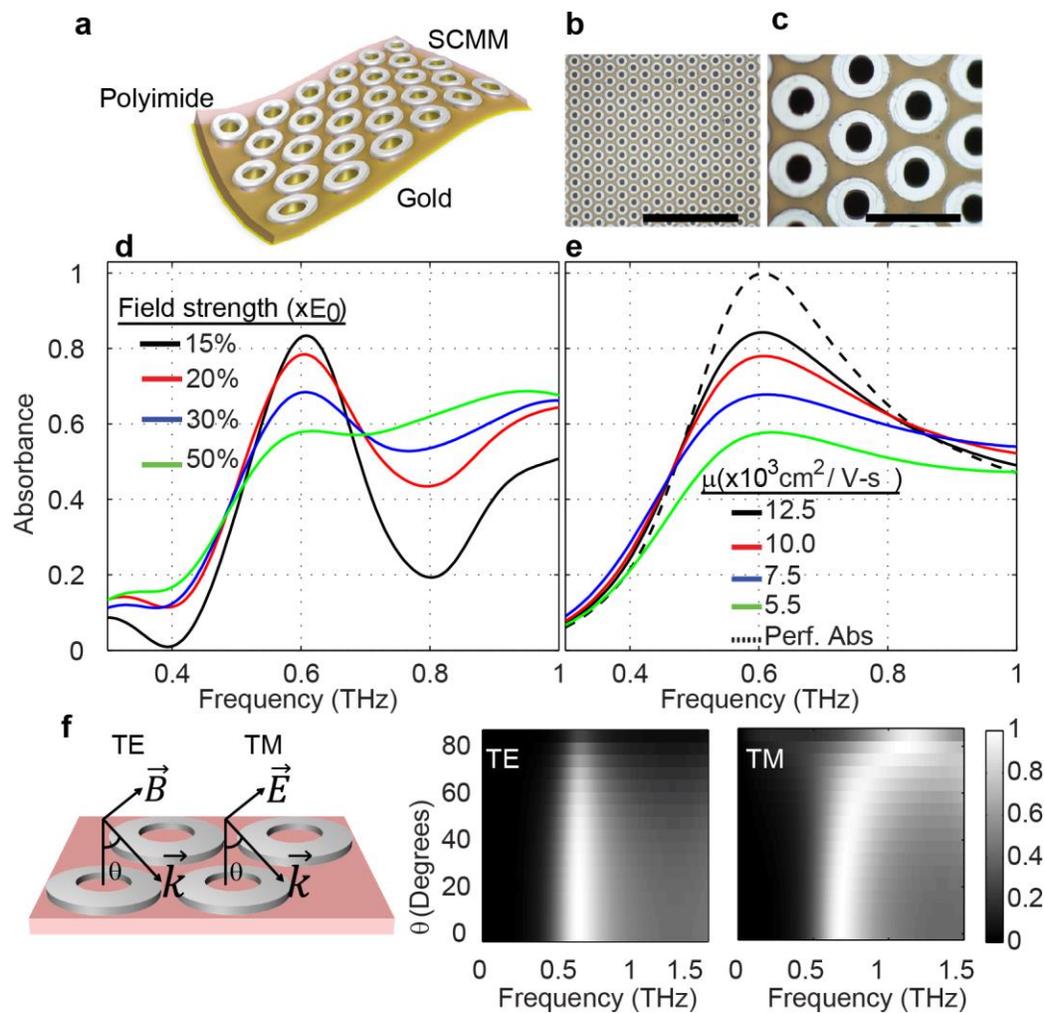

**Figure 4 | PSMM absorber on flexible substrate. a,** Representative sketch of the flexible semiconductor-based metamaterial absorber. **b,** Microscope image of the fabricated flexible absorber (scale bar is 500 µm). **c,** Close-up image of the rings. Rings have 30 µm outer and 15 µm inner radius on average with 72.7 µm hexagonal symmetry (scale bar is 100 µm). **d,** Terahertz time domain spectroscopy measurements showing the absorbance for increasing field strength. **e,** Simulated absorbance spectra for Drude models with varying mobility and effective mass. **f**, Simulated absorbance spectra of substrate-free absorber as a function of incidence angle (θ) for TE and TM polarized THz light.